\documentclass[pre,twocolumn,nofootinbib,superscriptaddress,showpacs,showkeys]{revtex4}

\usepackage{graphicx,epsfig}
\usepackage{amsfonts,amsmath,amssymb,amsthm,amscd}
\usepackage[T2A]{fontenc}
\usepackage{color}

\begin{document}

\title{Laser-heater assisted plasma channel formation in capillary discharge waveguides}

\author{N. A. Bobrova}
\affiliation{Institute of Theoretical and Experimental Physics, Moscow 117218, Russia}
\affiliation{Moscow Institute of Physics and Technology, Dolgoprudny, Russia}

\author{P. V. Sasorov}
\affiliation{Keldysh Institute of Applied Mathematics, Moscow, Russia}
\affiliation{Moscow Institute of Physics and Technology, Dolgoprudny, Russia}

\author{C. Benedetti}
\affiliation{Lawrence Berkeley National Laboratory, Berkeley, California 94720, USA}

\author{S. S. Bulanov}
\affiliation{University of California, Berkeley, California 94720, USA}

\author{C. G. R. Geddes}
\affiliation{Lawrence Berkeley National Laboratory, Berkeley, California 94720, USA}

\author{C. B. Schroeder}
\affiliation{Lawrence Berkeley National Laboratory, Berkeley, California 94720, USA}

\author{E. Esarey}
\affiliation{Lawrence Berkeley National Laboratory, Berkeley, California 94720, USA}

\author{W. P. Leemans}
\affiliation{University of California, Berkeley, California 94720, USA}
\affiliation{Lawrence Berkeley National Laboratory, Berkeley, California 94720, USA}

\begin{abstract}
A method of creating plasma channels with controllable depth and transverse profile for the guiding of short, 
high power laser pulses for efficient electron acceleration is proposed. The plasma channel produced by the hydrogen-filled capillary 
discharge waveguide is modified by a ns-scale laser pulse, which heats the electrons near the capillary axis. 
This interaction creates a deeper plasma channel within the capillary discharge that evolves on a ns-time scale, 
allowing laser beams with smaller spot sizes than would otherwise be possible in the unmodified capillary discharge.  
\end{abstract}

\maketitle

{\noindent}Extending the interaction length of an intense laser propagating in an underdense plasma by using a preformed channel 
is important for a wide variety of applications, including laser-plasma acceleration \cite{LWFA_review}, 
x-ray lasers \cite{Korobkin1996, Janulewicz2001}, harmonic generation \cite{Gaudiosi2006}, and direct 
laser acceleration \cite{York2008}.  Efficient laser-plasma accelerator (LPA) operation requires 
laser pulse propagation over a distance of the order of the acceleration length without significant 
diffraction.  Plasma channels, i.e., a plasma column with a minimum plasma density along the laser 
propagation axis, may be employed to optically guide short-pulse lasers, thereby extending the 
interaction length \cite{Esarey1997}.  Plasma channels may be created using a variety of techniques, 
including laser-induced hydrodynamic expansion \cite{Durfee&Milchberg1993,Durfee1995,Volfbeyn1999,Nikitin1999,Geddes2005}, 
ablative capillaries \cite{Ehrlich96,Kaganovich99,Kameshima2009}, fast Z-pinch assisted channel 
formation \cite{Hosokai2000}, and gas-filled discharge capillaries \cite{Butler2002,Luther2004,Gonsalves2007}.  
Hydrogen gas-filled discharge capillaries have been successfully used to generate electron beams 
in the GeV energy range \cite{Leemans,Karsch07,Rowlands08}.

Further increase in accelerated electron bunch energy lies in the utilization of more powerful lasers interacting 
with lower density plasmas. The BELLA (Berkeley Lab Laser Accelerator) project will employ a PW-class 
laser and up to a meter-scale plasma channel with on-axis density of about $10^{17}$ cm$^{-3}$ \cite{BELLA_AAC2010}, 
allowing 10 GeV electron bunch energy to be reached, and 
demonstrating the feasibility of laser plasma acceleration as a possible technology for the next generation 
of lepton colliders \cite{Schroeder2010}. However while the requirements of the maximal electron energy and available laser power 
fix the density and the spot size \cite{LWFA_review}, these can not be satisfied simultaneously by the available capillaries. 
Either the matched spot size is larger than required for laser guiding, or the small radius capillary, which would support 
guiding with the required matched spot, will experience extensive wall 
damage preventing the repetitive utilization.

In what follows we propose a robust method to guide short duration, high power laser pulses with the required spot 
size in a low density plasma channel, generated in a hydrogen-filled capillary discharge waveguide. After a parabolic density profile is 
established in the capillary by the discharge, a long (ns) laser pulse is fired along the capillary axis. The laser 
heats the electrons near the axis creating a secondary channel atop the already existing channel created by the 
discharge. The secondary channel width and depth can be controlled by varying the heating pulse duration and intensity, thus providing 
means of engineering a channel with the properties required for efficient acceleration of electrons.

In order to study the spatial and time evolution of the plasma electron density and temperature in a capillary discharge 
we performed computer simulations using a one-dimensional two temperature (ion and electron) single fluid magnetohydrodynamic (MHD) 
code \cite{Bobrova02}. The fact that the length of the capillary is much larger than its diameter justifies the 
utilization of the one-dimensional approximation, which is backed by the experimental results \cite{Ehrlich96} 
showing the distribution of the discharge parameters along the capillary axis to be uniform.

The physical model behind the MHD code was used to simulate the dynamics of dense Z-pinches~\cite{BRS} and both 
discharge-ablated~\cite{LPB} and pre-filled by gas capillary discharges \cite{Bobrova02}. We simulate 
the alumina capillary with an inner radius, $R_0$, of 500~$\mu$m, pre-filled with pure hydrogen of uniform temperature and 
density of $5.0\times 10^{-7}$~g/cm$^{3}$. The discharge is initiated by a pulse of current, which is approximately 
sinusoidal, with a half-period of $\sim 800$~ns and a peak value of 400~A, driven by an 
external circuit. The parameters of the capillary and the discharge are typical for the experiments with high 
power laser guiding. The plasma behavior is shown in Fig.~1. After the breakdown the current 
pulse heats the plasma and creates an azimuthal component of the magnetic field. However, pinching of 
the plasma is negligibly weak since the plasma pressure is much higher than the magnetic field pressure. 
The plasma is confined from expansion in the radial direction by the capillary walls.  We see three stages 
of plasma evolution. In the first stage ($<100$ ns), the radial distributions of the electron density, 
ion and electron temperatures 
and the degree of ionization remain homogeneous until the moment of almost full ionization of hydrogen. The 
electric current penetrates the plasma very quickly, and plasma is heated and ionized locally. The redistribution 
of plasma across the capillary channel begins at $\approx 100$~ns and lasts until $\approx 250$~ns. During 
this phase thermal conduction becomes significant. In the third stage the plasma remains 
almost immobile in the radial direction. The ion and electron temperatures are equal.
The radial distribution of plasma parameters corresponds to a quasi 
steady-state equilibrium at a given electric current. The plasma 
pressure is almost homogeneous. The  Ohmic heating is balanced mainly by thermal conduction to the relatively 
cold capillary wall. The temperature has its maximum on the axis and the hydrogen plasma is fully-ionized. 
Then, at constant pressure, a monotonic radially-decreasing temperature results in an axial minimum in the 
electron density profile of the plasma. During the interval 250~ns~$<t<$~800~ns the radial distribution of electron density is almost 
constant - at least to the extent that longitudinal plasma flow may be neglected. The time scale of the outflow 
is of the order of $L/c_s$, where $L$ is the length of the capillary, and $c_s=\sqrt{T_e/m_i}$ is the ion acoustic velocity. For $L=1$m the time of the longitudinal plasma outflow will be of the order of 30$\mu$s.

\begin{figure}[tbph]
\begin{center}
\includegraphics[width=9cm]{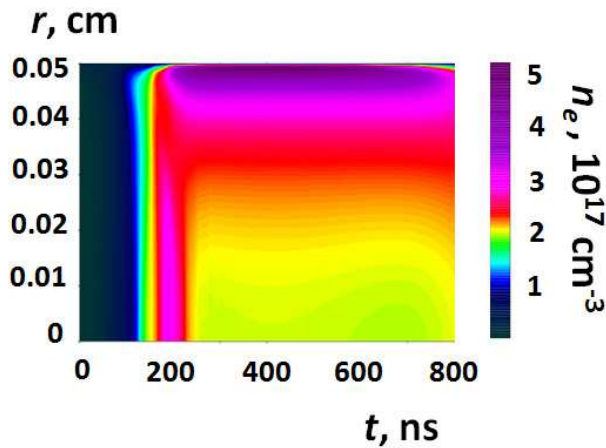}
\end{center}
\caption{Capillary discharge dynamics without laser heater: The distribution in the $(r,t)$ plane of electron density in
units of $10^{17}$ cm$^{-3}$. }
\end{figure}

For $t\gtrsim 250$ ns there is no screening of the axial electric field, and consequently it is 
uniform across the diameter of the capillary. There is no difference between electron  and ion temperatures. 
This allows us to use a simple model of the plasma under such equilibrium conditions, formulated in \cite{Bobrova02}. 
According to this model, a Gaussian beam will propagate through a plasma channel with a parabolic electron 
density profile of the form $n_{e}(r)=n_{e}(0)+n_{e}^{\prime \prime }(0)r^{2}/2$ with a matched spot size
\begin{equation}
W_{0}[{\rm \mu m}]\approx 1.48\times 10^5 \frac{\sqrt{R_{0}[{\rm \mu m}]}}{(zn_{i0}[{\rm cm}^{-3}])^{1/4}},
\label{Eq:18_54}
\end{equation}
where $R_0$ is the capillary radius, $z$ is the mean ion charge, and $n_{i0}$ is the initial ion density. For the 
case under consideration we find that $W_{0}\approx 145$~${\rm \mu m}$, which is typically too large for efficient 
LPA operation in the linear regime, \textit{i.e.}, to avoid strong laser evolution at high 
intensities (\textit{e.g.}, $60-90$ $\mu$m is required for BELLA experiment depending on the regime 
of operation \cite{BELLA_AAC2010}). In order to decrease the matched spot 
size it is necessary to use a capillary of a smaller radius. However, operation with small radii results in
to high wall lattice temperature and to melting or disruption of the wall material both by the 
discharge and the wings of the high power laser pulse which is to be guided through this capillary. 

\begin{figure}[tbph]
\begin{center}
\includegraphics[width=9cm]{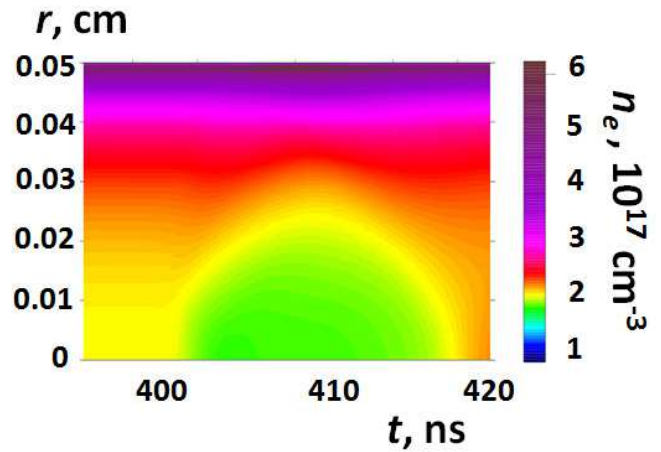}
\end{center}
\caption{Capillary discharge dynamics with laser heater: The distribution of the electron density in 
units of $10^{17}$ cm$^{-3}$ in the $(r,t)$ plane. }
\label{fig2}
\end{figure}

We propose to modify the plasma density profile in such a way that the capillary will gain the capability to guide 
pulses with smaller spot sizes, i.e., to provide additional heating of the capillary plasma near the axis by the 
absorption of the laser radiation. The heating laser pulse is fired along the capillary axis and is implemented 
into the MHD code in the following form:
\begin{equation}
P(r,t)=P_0 \cos^2 \left(\frac{\pi r}{2 R_{las}}\right)\, \, \sin^2 \left[\frac{\pi\, (t-t_i)}{\Delta t}\right]\, ,
\label{Eq:2}
\end{equation}
where $P(r,t)$ is the heating laser pulse power and is equal to 0 outside the intervals: $t_i<t<t_f$ and $r<R_{las}$. 
The constant coefficient $P_0$ is determined by the energy of laser pulse, $\mathcal{E}$, the parameter $R_{las}$ 
and duration $\Delta t =t_f-t_i$. The FWHM duration of the pulse is equal to $\Delta t/2$, and  
$R_{las}=\pi W_0/2\sqrt{2}$, which is chosen to match the form (\ref{Eq:2}) of the laser pulse to a 
Gaussian pulse with a spot size of $W_0$. We set the following 
parameters for the simulation: $\mathcal{E}=1$~J, $\Delta t/2=1$~ns and $R_{las}=161\, \mu$m. A one micron laser is assumed with critical electron density $n_{cr}=1.1\cdot 10^{21}$~cm$^{-3}$. The time of the maximum of electric current 
is chosen as the start of laser heating (t=400 ns). At that time the plasma is in equilibrium, hydrogen is fully 
ionized and there is a well established parabolic electron density profile to guide the heating laser pulse.  
The duration of the heating laser pulse ($\sim 1$ns) and the time of its propagation ($\sim 3$ns) in the capillary with length $L=1$m are much less than the hydrodynamic time ($\sim15$ns) at $t\approx 400$ns. Due to this separation of time scales, the energy deposition  of the laser radiation can be considered as homogeneous along the capillary. So we can consider a 1D (in radial dimension) approximation. We do not take into account the change of the matched spot size during the propagation of the heating laser pulse. The evolution of the radial distribution of electron density under the action of the heating laser pulse is plotted in Fig. 2. During the interval 401~ns~$<t<$~415~ns the radial distribution of electron density near the axis is different from the equilibrium one. An approximately parabolic electron density profile with smaller matched spot size exists over several nanoseconds.

\begin{figure}[tbph]
\includegraphics[width=9cm]{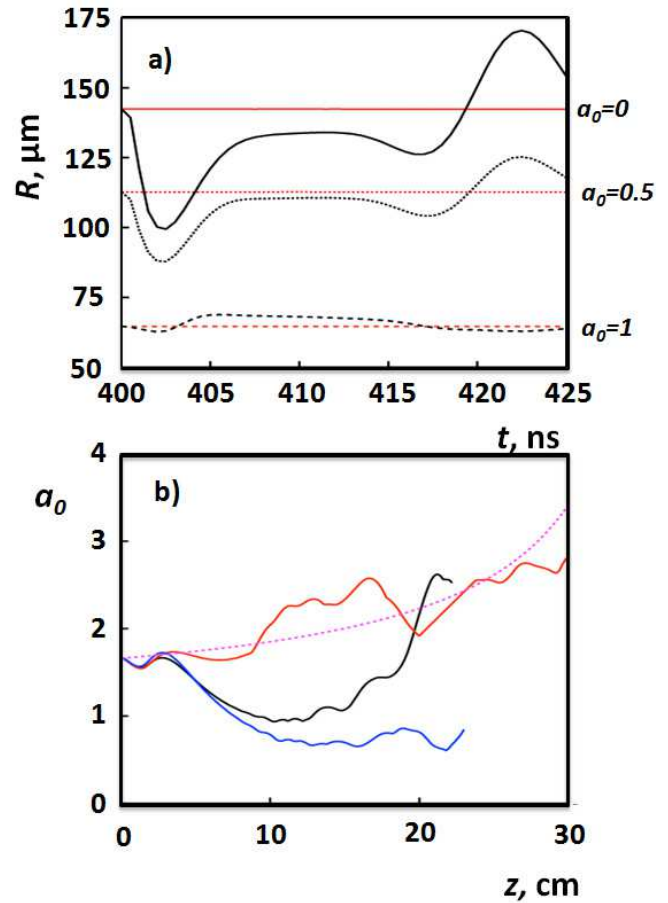}
\caption{
a) The dependence of the laser pulse matched spot size, $W_0$, on the delay between the guided pulse and the heating pulse for different values of $a_0=0,~0.5,$ and $1$, with (black curves) and without (red curves) laser heating.  
b) The dependence of $a_0$ on the propagation distance for the high power laser delay of 0 ns (black), 
2.5 ns (red), and 23.5 ns (blue). The dashed line corresponds to a 1D simulations. Here $a_0=1.7$. The simulations for blue and black curves were stopped at $z\approx 23$ cm, because in these two cases laser pulses show large mismatch oscillations.  
}
\label{fig3}
\end{figure}

We use the results of Ref. \cite{Benedetti} to calculate the matched spot size of the high power laser pulse, 
guided by a modified plasma density profile. In Fig. \ref{fig3}a the dependence of the matched spot size on 
the delay between the heating pulse and the high power laser pulse is shown. One can see that the matched spot size oscillates as the delay is increased and there are two distinct minima at $t\sim 402$ ns and at $t\sim 417$ ns. These oscillations are due to the fact that the cavity generated by the heating laser pulse has a form of a bubble (see Fig. 2), where the depth and radius change simultaneously as we move further away from the heating laser pulse along the bubble central axis. Since the matched spot size is the function of the channel depth, there are two points along the central axis that correspond to the matching of the laser pulse to the modified plasma density profile. In Fig. 3a these two points correspond to the local minima in $R=R(t)$. We use the first minimum to guide the high intensity laser pulse, since it supports guiding of smaller matched spot size laser pulses. Although $R$ changes as a function of time near the minimum matched spot size, on a nanosecond scale the variation is sufficiently small to significantly affect the guiding. The ns level of synchronization between long and short pulses was demonstrated in experiments \cite{fs+ns,fs+ns1}.

The propagation of the laser pulse with the anticipated parameters of BELLA laser (40 J, 100 fs and $W_0=64$ $\mu$m) was modeled with the code INF\&RNO (INtegrated Fluid \& paRticle simulatioN cOde), which is a 2D cylindrical code that solves the full wave equation for the laser envelope evolution coupled to the equation of motion for the plasma (PIC and cold fluid descriptions are available) and to Maxwell equations for the electromagnetic field in the plasma. The plasma responds to the laser field via the averaged laser ponderomotive force. The adoption of the cylindrical geometry allows the description of 3D physics (laser evolution, electromagnetic field structure) at 2D computational cost. The code features an improved laser envelope solver which enables an accurate description of the laser pulse evolution deep into depletion even at a reasonably low resolution. Additional details on INF\&RNO may be found in Ref. \cite{INFERNO}.

The results are shown in Fig. \ref{fig3}b, where the dependence of $a_0$, the peak value of the dimensionless vector potential of the laser electromagnetic field, on the propagation distance for the plasma density profiles at 
400 ns, 402.5 ns, and 423.5 ns is presented.  The results of 1D simulations (the dashed curve) demonstrate 
that the slow increase of $a_0$ is due to the longitudinal pulse evolution and not to the transverse dynamics. 
From the close matching of the 1D $a_0$ evolution to the 2D result in red in Fig. 3b one can see that modified 
plasma density profile provides matched guiding of a high power laser pulse with a 1.5 ns delay after the 
peak of the heating pulse. This delay corresponds to the minimum matched spot size that can be achieved with this 
modified density profile. The depth and width of the modified density profile can be controlled by the energy and 
duration of the heating laser pulse, making it possible the matching of different high power laser spot 
sizes for efficient guiding.   

We notice that the dependence of the peak value of $a_0$ on the propagation distance demonstrates oscillatory behavior. The reason for the oscillations is connected with the fact that it is not possible to ensure the matching to the plasma density profile uniformly along the high power laser pulse. The oscillations of the non-matched parts of the laser pulse as well as their interaction between themselves and with the matched part result in these non-regular oscillations, seen in Fig. 3b.        

In conclusion, we have proposed a robust method of controllable modification of the plasma density profile in the 
gas-filled capillary discharge waveguide. It is achieved by heating the plasma near the capillary axis by a 
long (ns scale) laser pulse. The laser heating results in the generation of a secondary plasma channel atop 
of the existing discharge-generated channel, which serves the purpose of guiding high power laser pulses with 
smaller spot sizes than the non-modified plasma density profile can provide. We demonstrated through MHD 
modeling that such profiles can be generated and have shown that they should be capable of guiding high power lasers. This method enables larger capillary radii, thereby preventing the extensive wall damage and thus greatly 
increases the lifetime of the capillary. This is extremely important for the LPA operation in the high repetition regime.
The multidimensional effects (including laser heater evolution), neglected in the present paper, will be a subject of future studies for different capillary geometries.  

We appreciate support from the NSF under Grant No. PHY-0935197 and the Office of Science of the US DOE under 
Contract No. DE-AC02-05CH11231. N.B. acknowledges support from DOE that allowed her to visit LBNL to work on the 
laser heating method proposed by W.P.L.

%\pagebreak

\end{document}